# "Comparing university organizational units and scientific co-authorship communities"

Uwe Obermeier [a], Michael J. Barber [b], Andreas Krueger [a], Hannes Brauckmann[a]

**[a], University College Dublin**
**UCD Innovation Research Unit (IRU), University**
**8 Belfield Office Park, Beaver Row, Clonskeagh, Dublin 4, Ireland**

**[b] AIT Austrian Institute of Technology GmbH**
**Department Foresight & Policy Development**
**TechGate Vienna, Donau-City-Straße 1, 1220 Wien, Austria**

**Abstract:**

A co-authorship network of scientists at a university is an archetypical example of a complex evolving network. Collaborative R&D networks are self-organized products of partner choice between scientists. Modern science is, due to the immanent imperative of newness, strongly interdisciplinary. Crossovers between the different scientific disciplines and organizational units are observable on a daily basis. Since collaborative research has become the dominant and most promising way to produce high-quality output, collaboration structures are also a target for research and management design.

We study co-authorship covered by the Citation Index within University College Dublin, a large Irish university. We focus especially on collaborations between organizational units, such as schools or colleges. We compare the centrality of the brokerage individuals within their organizational units. The network of co-authorship is analyzed with standard network measures, with typical features for complex networks observed. We use the bipartite network of authors and papers for the identification of communities of collaborators and clusters of papers. The observed communities are compared to the formal organizational units of the university.

## 1. Introduction

A co-authorship network of scientists at a university is an archetypical example of a complex evolving network. Collaborative R&D networks are self-organized products of partner choice between scientists. Various theoretical frameworks have been used to describe recent academic research: Though these frameworks differ in many respects, they all point to a more collaborative R&D "network mode" of knowledge production. Concepts such as "Mode 2" (Gibbons et al., 1994), "Academic Capitalism" (Slaughter and Leslie, 1997) , "Post-Academic Science" (Ziman, 2000), or the "Triple Helix" (Etzkowitz and Leydesdorff, 2000) not only refer to external collaborations of universities with industry, government and other actors, but report changed practices inside academia.

In universities, the isolated researcher in the ivory tower has been widely replaced by interdisciplinary teams in collaborative research projects (Wuchty et al, 2007). Crossovers and co-operations between different scientific disciplines, different organisational units, and external actors are a common and increasing phenomena of academic reality (Guimerà et al, 2005). Since collaborative research has become the dominant and most promising way to produce high-quality output (Jones et al, 2008), collaboration structures are also a target for research and management design (Bozeman, Lee 2005). Collaborative research projects, co-authored publications, or multidisciplinary excellence networks in universities point to the peer network mode of today´s knowledge production.

Our study investigates co-publications between different organisational units of University College Dublin (UCD). Using networks inferred from co-publication data, we attempt to infer the interdisciplinary publication culture inside UCD. The combination of co-authorship and SNA is often used for mapping, analysing, and evaluating scientific research activity (Olmeda-Gómez et al 2008, Barabasi et al 2002, Yousefi-Nooraie et al 2008).

The study is set up as an academic research project but is intended to contribute to the self-monitoring mechanisms of UCD as well. We assume that demand from academic management for consultancy will grow due to a rising need for legitimacy. Therefore it is important to embed network analysis in a methodological and ethical-regulatory context to evaluate its applicability.

## 2. Data, Method, Results

Data for the present study reflects the publication output of permanent academic staff at UCD during the period 1998-2007, drawn from the ISI Thompson Citation Index (Web of Knowledge). We include publications from the Science Citation Index (SCI), Social Science Citation Index (SSCI), and the Arts and Humanities Index (A&HCI), which are classified as normal articles, reviews, letters and notes.

At UCD, all members of academic staff belong to Schools or Research Units, which are themselves contained within a College. For the purposes of our study, we measure interdisciplinarity as co-publication activity of UCD authors belonging to different Colleges and Schools.

### Researchers with publications within and between colleges 1998-2007

Table 1 shows researchers with at least one publication covered by the Citation Index. We can observe within the *College of Arts & Celtic Studies* (1) there are 62 researchers publishing covered by the SCI. The *College of Business & Law* (2) has 44 researchers, the *College of Engineering, Mathematical & Physical Sciences* (3) has 139 researchers, and the *College of Human Sciences* (4) has 97 researchers. The highest number of (SCI-) publishing researchers comes from the *College of Life Sciences* (5), which is represented by 291 researchers.

| College | 1 | 2 | 3 | 4 | 5 | 6 |
|---|---|---|---|---|---|---|
| Count | 62 | 44 | 139 | 97 | 291 | 9 |

Table 1: (1) College of Arts & Celtic Studies, (2) College of Business & Law, (3) College of Engineering, Mathematical & Physical Sciences, (4) College of Human Sciences, (5) College of Life Sciences, (6) Others, e.g., Finance Office. Source: UCD Research

**Co-authored publications within and between different colleges 1998-2007**

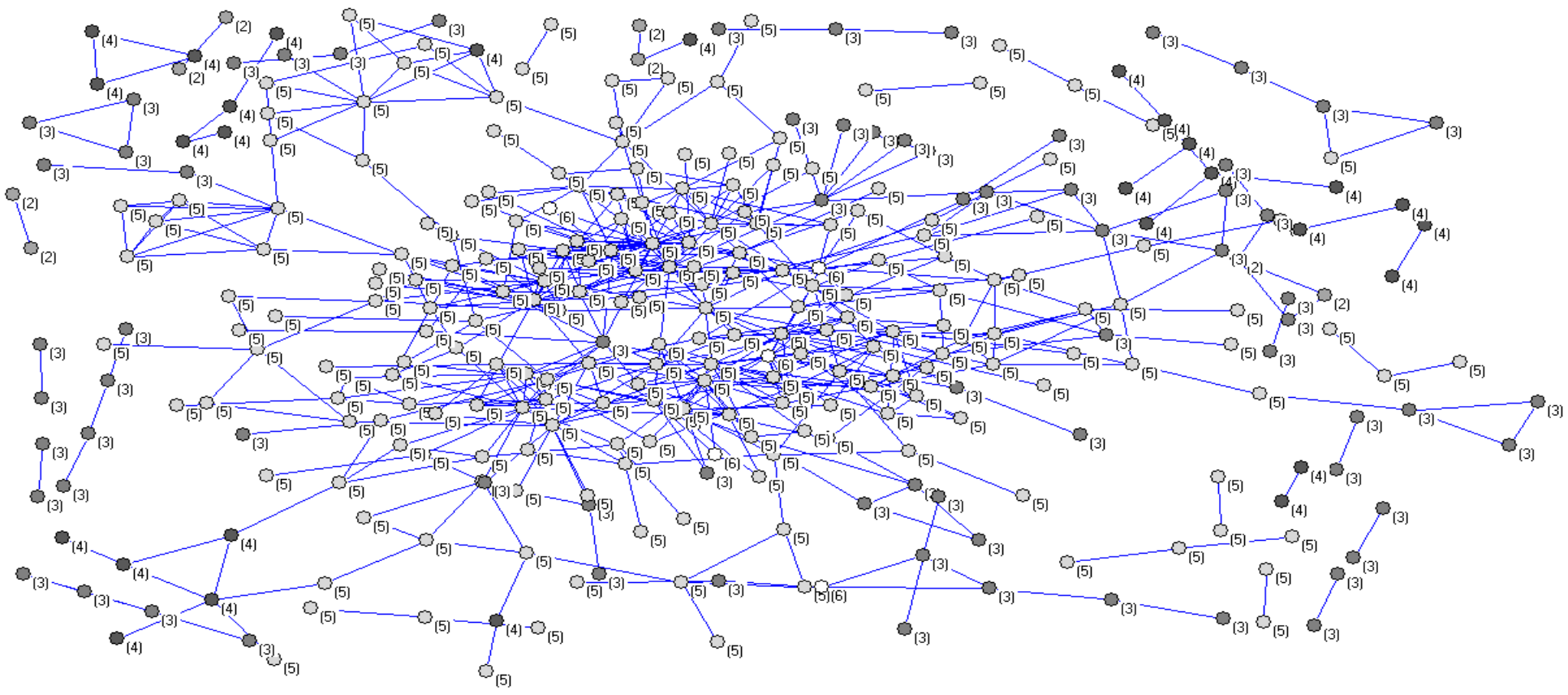

Figure 1: 1997-2007 Nodes (2) College of Business & Law ,Nodes (3) College of Engineering, Mathematical & Physical Sciences, Nodes (4) College of Human Sciences, Nodes (5) College of Life Sciences, Nodes (6) others. Data Source: UCD Research

| College | 2 | 3 | 4 | 5 | 6 |
|---|---|---|---|---|---|
| Count | 10 (23%) | 74 (53%) | 30 (31%) | 222 (76%) | 5 |

Table 2: College, researchers with co-authorship within UCD, % of researchers represented in the SCI from overall

In Figure 1, we only focus on researchers with co-authored publications within UCD. Every node represents a researcher with at least one co-authored publication; a link between two nodes represents at least one co-published paper (for characteristics of the networks see appendix). We can observe that within the *College of Business & Law* (2), 23% of researchers are connected by co-authorship. Within the *College of Human Sciences*, (4), we observe that 31% of researchers are connected by co-authorship. The highest level of co-authorship between researchers within UCD we can observe is within the *College of Engineering, Mathematical & Physical Sciences,*(3), and the *College of Life Sciences,* (5), with 53%; and 76% of the overall representation in SCI respectively. Some individuals like Heads of Schools or Research Management were categorized separately.

| College | 2 | 3 | 4 | 5 | 6 |
|---|---|---|---|---|---|
| Count | 4 (40%) | 25 (34%) | 5 (17%) | 61 (27%) | 5 |

Table 3: Authors with co-authorship outside their own college, Percentage of Table 2 researchers

We can observe a high level of co-authorship from authors outside their own college from the *College of Engineering, Mathematical & Physical Sciences*, which has 34% of the connected researchers at UCD, and the *College of Life Sciences,* which has 27%.

In Table 4 and Figure 2 we focus on co-authorship between different schools to investigate interdisciplinarity (within Figure 2 there are also links with co-authorship less than 10 included). The highest level of co-authorship between different Schools we can observe is within the Schools of the *College of Life Sciences*. Some very central Schools are *Medicine and Medical Sciences*, *School of Biomolecular & Biomedical Science* and *School of Agriculture, Food Science & Veterinary Medicine*.

**Interdisciplinary Publications between different schools 1998-2007**

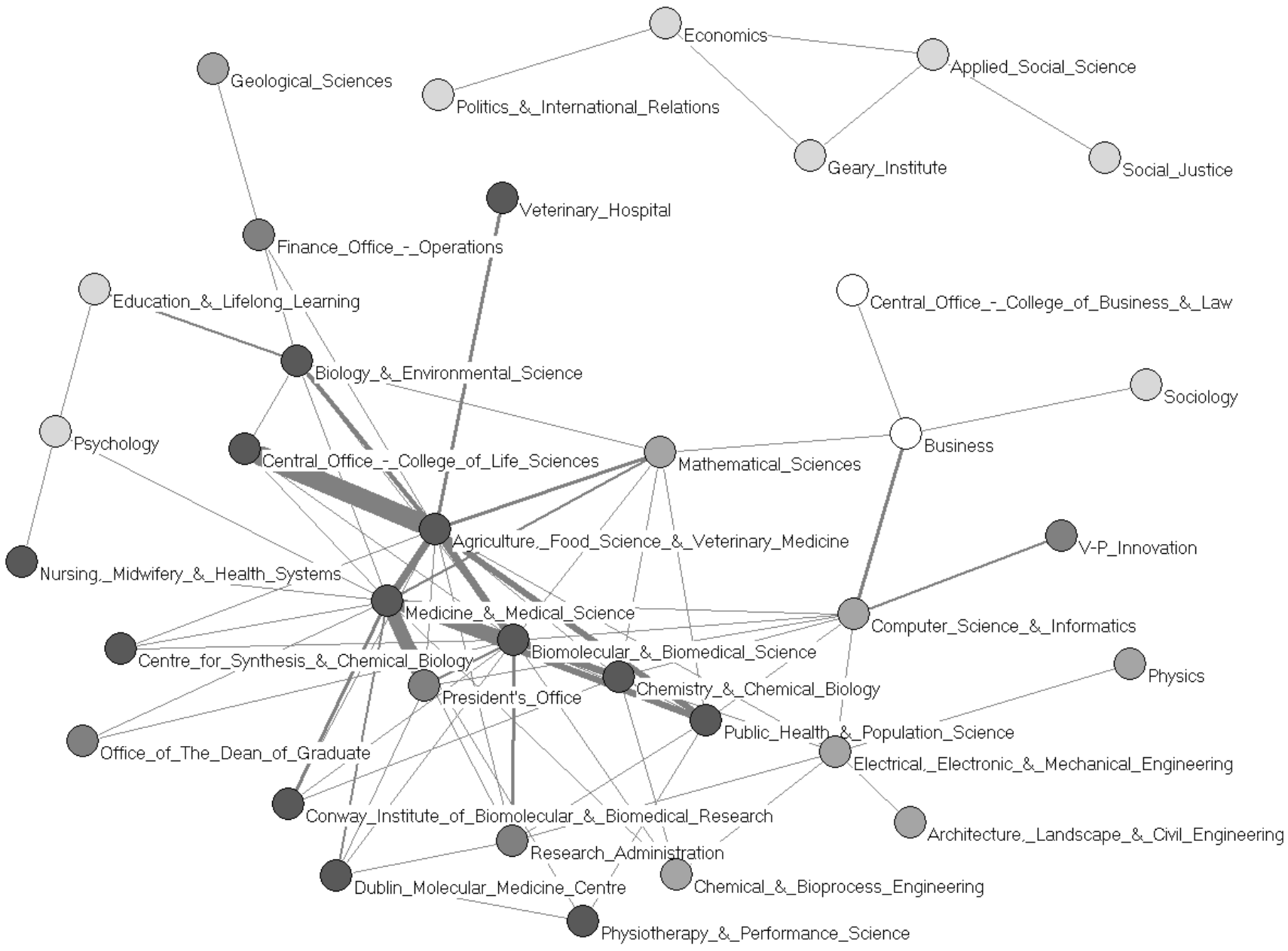

Figure 2 1997-2007 Data Source: UCD Research.

There is a high level of co-authorship between the *School of Mathematical Sciences* and the *School of Agriculture, Food Science & Veterinary Medicine*, and the *School of Medicine and Medical Science*. There are also a large number of co-authored papers between the *School of Computer Science & Informatics* and the *School of Business.*

| Co-Auth | School 1 Name | School 2 Name |
|---|---|---|
| 100 | Central Office - College of Life Sciences | Agriculture, Food Science & Veterinary Medicine |
| 71 | Medicine & Medical Science | President's Office |
| 64 | Medicine & Medical Science | Biomolecular & Biomedical Science |
| 40 | Biomolecular & Biomedical Science | Agriculture, Food Science & Veterinary Medicine |
| 39 | Chemistry & Chemical Biology | Biomolecular & Biomedical Science |
| 38 | Medicine & Medical Science | Agriculture, Food Science & Veterinary Medicine |
| 36 | Public Health & Population Science | Agriculture, Food Science & Veterinary Medicine |
| 34 | Public Health & Population Science | Medicine & Medical Science |
| 21 | Biology & Environmental Science | Agriculture, Food Science & Veterinary Medicine |
| 18 | Veterinary Hospital | Agriculture, Food Science & Veterinary Medicine |
| 16 | Mathematical Sciences | Agriculture, Food Science & Veterinary Medicine |
| 15 | Conway Institute of Biomolecular & Biomedical | Medicine & Medical Science |
| 15 | Research Administration | Biomolecular & Biomedical Science |
| 14 | Computer Science & Informatics | Business |
| 13 | President's Office | Biomolecular & Biomedical Science |
| 12 | Dublin Molecular Medicine Centre | Medicine & Medical Science |
| 10 | Mathematical Sciences | Medicine & Medical Science |
| 10 | V-P Innovation | Computer Science & Informatics |

Table 4: Frequency of co-authorship between different schools, at least 10 co-authored papers

Using co-authorship as an indicator for mapping the interdisciplinary cooperation structure of the university shows that within and between single Colleges the co-publication activity differs significantly. We observe for example:

- There are no co-authored publications with or within the *College of Arts & Celtic Studies*, at least not as covered by the Citation Index.
- The largest amount of intra-College co-authorship we can observe is within the *College of Life Sciences*.
- The amount of co-publications between different Schools of a particular College points to working interdisciplinary cooperation structures between thematically close disciplines (specifically, the Schools of Medicine & Medical Sciences, Biomolecular & Biomedical Science, and Agriculture, Food Science & Veterinary Medicine frequently co-publish with other Schools of their College)

- The *College of Life Sciences* also shows the highest number of interdisciplinary co-publications with other Colleges, especially with the *College of Engineering, Mathematical & Physical Sciences*, which points to working interdisciplinary cooperation structures between fields that are more thematically separated.
- This intra-College cooperation is concentrated in particular Schools, i.e.
  - the *School of Agriculture, Food Science & Veterinary Medicine* and the *School of Medicine & Medical Science* (both College of Life Sciences) frequently co-publish with the *School of Mathematical Sciences* (College of Engineering)
  - or the *School of Computer Science & Informatics* (College of Engineering) and the *School of Business* (College of Business and Law) are frequent co-publishers.

These observations lead us to assume that there are specific profiles of Schools and individuals with skills crucial to connect and contribute to other Schools/individuals. This ranges from cooperation between thematically close areas to bridging huge disciplinary distances. We can easily identify "champions" of interdisciplinary co-publication, brokers between disciplines, and areas where this is not working at all.

**What is the position of the brokerage individuals?**

From these results, we next investigate the positions of brokerage individuals, the people connecting the different organizational units. Are they the central ones within their own organizational units or schools? Is there any relation to the thematic fields and, if so, what does it look like?

Heinze et al (2009) distinguishes two main lines of thought. Network brokerage argues that people who are placed at the intersection of heterogeneous social groups have an increased likelihood of drawing upon multiple knowledge sources, leading to the generation of new ideas (Burt, 2004). In contrast, proponents of cohesive collaborative networks argue for the benefits of trust, shared risk taking and easy mobilization in facilitating information and knowledge transfer; according to these studies, individuals with cohesive ties are likely to be involved in innovations (Uzzi and Spiro 2005, Gloor 2006).

Above, we analyzed the schools strong in inter-organizational co-authorship. We focused especially on the schools from the *College of Life Sciences*; the schools from the *College of Engineering, Mathematical & Physical Sciences*; and the *School of Business* from the *College of Business and Law*.

We now consider the positioning of authors from the six schools. We distinguish the edges of the network as being intra-school, where authors are from the same school, or inter-school, where authors are from different schools. Authors with strong intra-school links are taken to be central to their own school, while those with strong inter-school links are taken to be central to the overall network. These centralities correspond to their degree centralities in two auxiliary networks made by segregating the two types of links. To assess the centrality values, we calculated Pearson's Correlation Coefficient and Spearman's Rank Correlation.

| School | Nr of vertices | Pearson Correlation Coefficient | Spearman Rank Correlation |
|---|---|---|---|
| Agriculture, Food Science & Veterinary Med | 64 | 0.43 | 0.34 |
| Computer Science & Informatics | 16 | -0.13 | -0.15 |
| Mathematical Sciences | 13 | -0.38 | -0.34 |
| Medicine & Medical Science | 68 | 0.39 | 0.40 |
| Business | 9 | -0.35 | -0.64 |
| Chemistry & Chemical Biology | 9 | 0.59 | 0.27 |

Table 5: Comparison of degree centrality of individuals within selected schools to overall network.

The two correlation measures are consistent as to which correlations are positive and which are negative. Positive correlations are observed for the schools strongly related to the *College of Life Sciences*, while the others show negative correlations. Bearing in mind that we have a relatively small data set from a single university, these results indicate:

- Within thematically close Schools, the correlation is positive. Brokerage individuals are more central within their own units.

- Between thematically distant Schools, the correlation is negative. Brokerage individuals are less central within their own units. Individuals central in their own school are seldom brokers.

**Organizational units and co-authorship communities**

To further investigate the role that the UCD colleges and schools play in determining co-authorship, we infer collaborative communities from the structure of the network itself. We focus our attention on the bipartite network of authors and publications formed by linking each publication to its author. This network is not connected; we restrict investigation to the largest component, containing 234 authors.

The identification of communities from network structure is a topic of great recent interest. Formulation of the problem presents two main challenges. First, the notion of community is imprecise, requiring a definition to be provided for what constitutes a community. Second, community solutions must also be practically realizable for networks of interest. The interplay between these challenges allows a variety of community definitions and community identification algorithms suited to networks of different sizes, as measured by the number of vertices *n* or edges *m* in the network (for useful overviews, see Newman 2004b, Danon et al. 2005, and Fortunato and Castellano 2008).

A prominent formulation of the community-identification problem is based on the modularity *Q* introduced by Newman and Girvan (2004). The quality of communities given by a partition of the network vertices is assessed by comparing the number of edges between vertices in the same community *c* to the number expected from a null model network. Good quality communities then have more intra-community links than would be expected from the model, and fewer inter-community links than expected.

As we consider a bipartite network of authors and publications, we use a bipartite null model (Barber 2007). Formally, this gives

$$Q = \frac{1}{m}\sum_{c}\sum_{i,j\in c}\left(A_{ij} - \frac{k_i k_j}{m}\right)$$

where the $A_{ij}$ are components of the adjacency matrix for the network, the $k_i$ are the degrees of the network vertices, and the sums over *i* and *j* are restricted to run over the authors and publications, respectively.

Community identification is then a search for high modularity partitions of the vertices into disjoint sets. An exhaustive search for the globally optimal solution is only feasible for the smallest networks, as the number of possible partitions of the vertices grows far too rapidly with network size. Several heuristics exist to find high-quality, if suboptimal, solutions in a reasonable length of time. Here, we use a two-stage search procedure:

1. Agglomerative hierarchical clustering, where small communities are successively joined into larger ones such that the modularity increases. This stage is based on the so-called fast modularity (FM) algorithm (Clauset et al. 2004), adapted to work with bipartite networks.
2. Greedy search, where vertices are moved amongst existing communities to ensure the resulting partition is at a local optimum of modularity. This stage uses the bipartite, recursively induced modules (BRIM) algorithm (Barber 2007).

The coarse structure is found with FM, with incremental improvements provided by BRIM.

Using the above search procedure, we identify communities of authors and publications for UCD. The community partition found has modularity *Q* = 0.93 (compared to a maximum of *Q* = 1), indicating a clear community structure in the network. There are 53 communities, in contrast to the 5 colleges and 26 schools corresponding to authors present in the network component. In Figure 3, we show the distribution of community sizes, as measured by the numbers of authors in the communities.

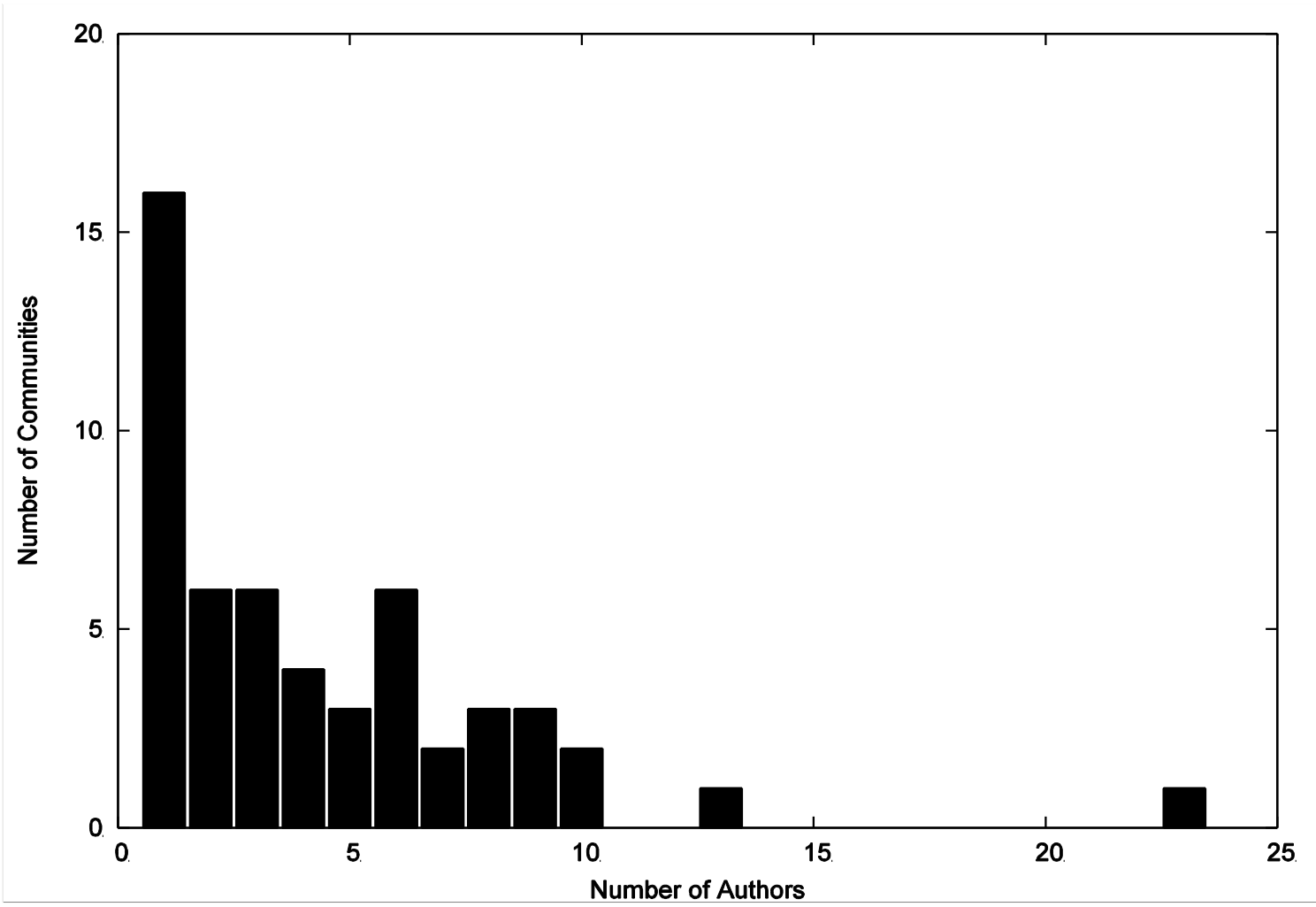

Figure 3. Frequency of community size, as measured by the number of authors in the communities. The 53 communities range from having just a single author to as many as 23 authors.

To further compare the structurally identified network communities to the organization in terms of colleges or schools, we use the normalized mutual information $I_{norm}$ (Danon et al. 2005). The normalized mutual information allows us to measure the amount of information common to two different partitioning schemes. We explain it for the schools; the same explanation holds, *mutatis mutandis*, for the colleges. We take one of the partitions to be authors in the found communities, and the other to be a partition defined by the schools. The value of $I_{norm}$ increases monotonically from zero to one as the correlation between the two partitions increases. At the extremes, we have $I_{norm} = 1$ when the found communities match the school-based communities and $I_{norm} = 0$ when they are independent.

Comparing the colleges to the structurally identified communities produces $I_{norm} = 0.20$, indicating that the colleges reveal little about scientific collaborations. The comparison for schools gives $I_{norm} = 0.56$, indicating that the schools partially, but by no means completely, describe the collaborative structures at UCD.

To clarify the limitations of the explanative power for the schools, we consider the diversity of the authors in the communities. In Figure 4, we show how many communities have authors from different numbers of schools. Most common is that all authors in a community are from the same schools, but this includes the 16 communities which contain only one author. Excluding these, nearly 70% of the multi-author communities involve inter-school collaboration, with several involving authors from five different schools. Partitioning the authors based on the schools alone cannot account for this, indicating that neither the colleges nor the schools provide a complete understanding of the scientific collaborations at UCD. This is consistent with the extensive inter-college and inter-school collaboration described above.

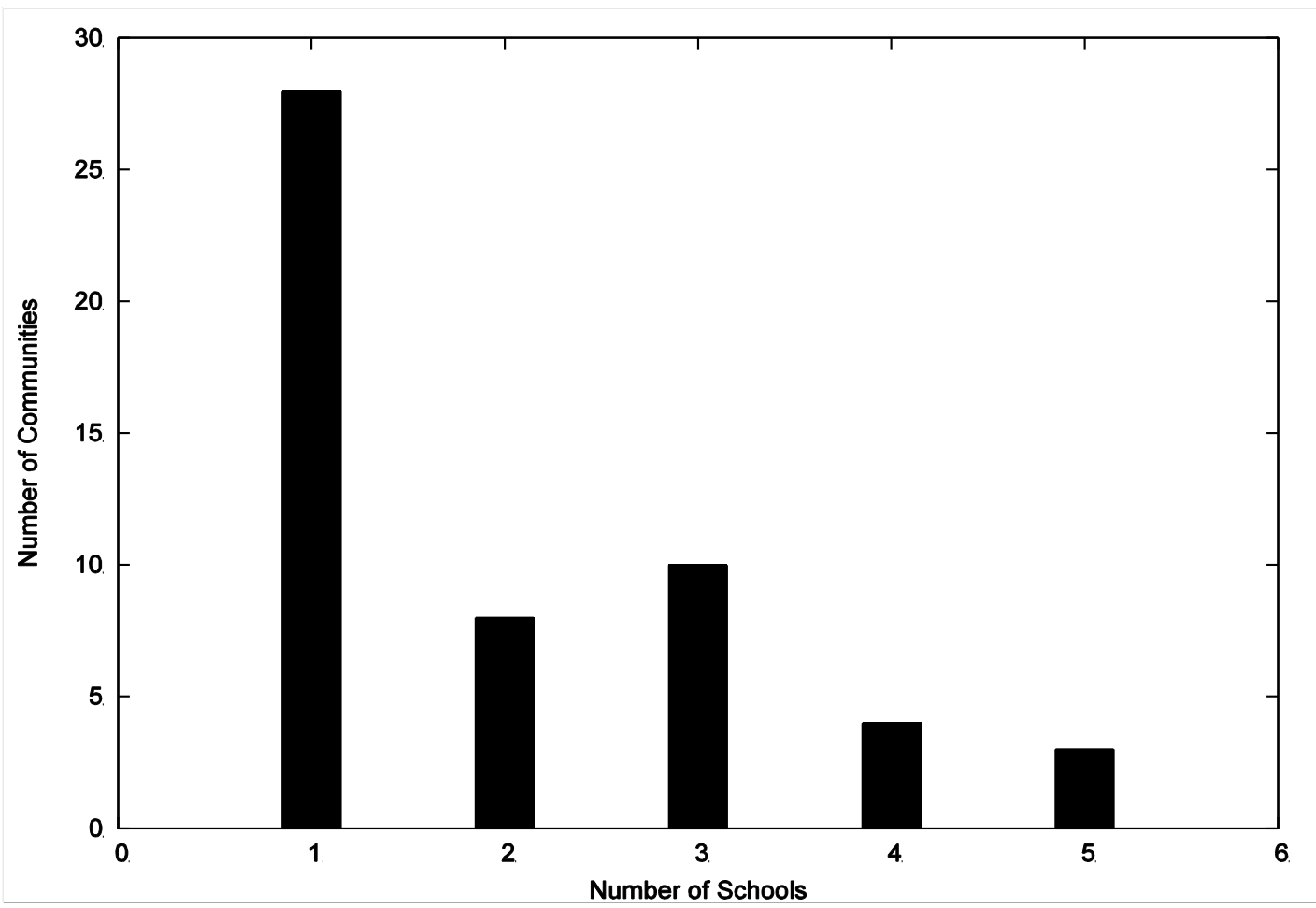

Figure 4. Diversity of the authors in communities can be assessed by examining the number of schools found in each community. Nearly half of all communities involve cross-school co-authorship. Note that the 28 single-school communities include all 16 of the single-author communities.

**Inter-school collaborations at the community level**

The nature of the inter-school collaborations at the community level is also of interest. We should expect communities spanning two or more schools to reflect interdisciplinary research at UCD, and conversely to indicate which schools are most important for interdisciplinary work. To quantify this, we calculate the number of multi-school communities in which each school is found; communities where all authors are from the same school are not considered. In Table 6, we show the schools best represented in the communities, along with the count of multi-school communities in which they appear. We exclude the five schools appearing in just two communities and the nine schools appearing in just a single community. From this community-based viewpoint, we again see that schools focused on the life sciences are central to inter-school work.

| Communities | School |
|---|---|
| 12 | Medicine & Medical Science |
| 11 | Agriculture, Food Science & Veterinary Medicine |
| 9 | Biology & Environmental Science |
| 8 | Biomolecular & Biomedical Science |
| 3 | Electrical, Electronic & Mechanical Engineering |
| 3 | Nursing, Midwifery & Health Systems |
| 3 | Conway Institute of Biomolecular & Biomedical Research |
| 3 | Chemistry & Chemical Biology |
| 3 | Mathematical Sciences |
| 3 | Public Health & Population Science |

Table 6: Number of multi-school communities by school

Extending the above idea, we also examine the role of particular collaborations between schools in the multi-school communities. This consists of counting the number of communities where a particular pair of distinct schools is present. Values for the most frequent pairs are shown in Table 7, excluding the fourteen school pairs appearing in just two communities and the 46 community pairs appearing in just a single community. Here, the schools are all from the life sciences, with the sole exception of the *School of Electrical, Electronic & Mechanical Engineering*, lending further support to the notion that schools from the *College of Life Sciences* are central to inter-school work.

| Communities | School | School |
|---|---|---|
| 5 | Agriculture, Food Science & Veterinary Med | Biology & Environmental Science |
| 4 | Biomolecular & Biomedical Science | Medicine & Medical Science |
| 3 | Agriculture, Food Science & Veterinary Med | Electrical, Electronic & Mechanical Eng. |
| 3 | Agriculture, Food Science & Veterinary Med | Medicine & Medical Science |
| 3 | Biology & Environmental Science | Medicine & Medical Science |

Table 7. Number of communities for inter-school collaborations.

## 3. Discussion

We analyzed a co-authorship network of scientists at a university as an archetypical example of a complex evolving network. The distribution of the number of publications indexed by the Science Citation Index is not uniform across schools, with the greatest number of publications found in the *College of Engineering, Mathematical & Physical Sciences* and the *College of Life Sciences*. These two colleges are also the colleges with the highest percentage of researchers represented with co-authorship within the university: the *College of Life Sciences* with 76% and the *College of Engineering, Mathematical & Physical Sciences* with 53%.

Inter-school co-authorship is most frequently observed between schools in the *College of Life Sciences*.

Inter-school publications are most frequently observed with authors from the *School of Agriculture, Food Science & Veterinary Medicine* and from the *School of Medicine & Medical Science* (both College of Life Sciences); interestingly, inter-school co-authorship between these two schools is not the most common. Schools frequently co-publishing across colleges, and presumably at greater thematic distance, are the *School of Mathematical Sciences* (College of Engineering) with the Schools from *College of Life Sciences* or the *School of Computer Science & Informatics* (College of Engineering) and the *School of Business* (College of Business and Law).

Further we analysed the positions of the brokerage individuals. We found a positive correlation of centrality for the brokerage individuals for thematically close schools, e.g., the *School of Agriculture, Food Science & Veterinary Medicine*, the *School of Medicine & Medical Science* as well for the *School of Chemistry & Chemical Biology*. In contrast we found a negative correlation of centrality for thematically distant schools like Computer Science & Informatics, Business and Mathematics.

Finally, we investigated co-authorship using author communities identified from the structure of the co-authorship network. The UCD colleges provide little insight into the author communities. The schools correspond better to the communities, but fail to provide a complete picture due to communities featuring inter-school collaboration. The schools involved in multi-school communities are most frequently from the life sciences, consistent with what we found by starting with the organizational structures at UCD.

Scientific disciplines are differently represented in (co-)authorship databases such as the ISI Thompson Citation Index. This biases our data and, accordingly, our results. To interpret these results appropriately, we need to relate them to the different publication cultures in the scientific disciplines.

These differences manifest themselves in various aspects, e.g. (co-)authorship is less frequent in theoretical than in experimental fields. We illustrate this point using disciplinary differences in publication culture: van Raan (2005) estimates that in the medical sciences the share of publications covered by the Citation Index is about 80 to 95 percent; in the social sciences it is much lower, at 50-70 per cent coverage in psychology; and in the engineering fields, it is about 50 per cent.

From this, we know that our results are biased. There is much (co-)publication activity which is not recorded in the database and, accordingly, does not appear in our results. Further, Laudel (2002) shows that co-publication as an indicator does not cover important aspects of research collaboration: most collaboration within a university never results in a co-authored publication. Thus, even if the data were to be weighted against scientific publication cultures, co-publication would need to be complemented by other indicators of interdisciplinary cooperation dynamics. Laudel and Glaeser (2006) have presented in detail how and where scientific practice of research collaborations slips through the nets of current evaluation instruments. This especially refers to using publications as the only measure for assessing research collaboration and performance within an over-simplified and wrongly standardized evaluation approach.

To avoid these pitfalls, we must discuss the first results of our study in a methodological and ethical-regulatory context, especially when communicating them to university management. The methodological context concerns the discipline- and organization-dependent usability of bibliometric approaches such as (co-)publications. The ethical-regulatory context refers to requirements of protecting personal data and ensuring transparent management procedures. Only by taking into account the methodological context of Social Network Analysis based higher education research, can university management practice begin to translate these findings into strategic options for university development.

**Appendix: Author Graph and Publication Graph**

We calculated standard network measurements for the UCD co-authorship network. We considered two projection networks from the bipartite author-publication network. These are the author projection, where authors are linked if they have co-authored a publication, and the publication projection, where publications are linked if they share a common author.

In Table A1, we show properties of the two projections. Note that these classical SNA measurements (Wasserman 1994) are based on *unweighted projections*, meaning that, e.g., two authors are linked if they share publications, but the actual number of common publications is irrelevant. In Table A2 we summarize properties of distributions derived from the network edges. The number of authors per publication and the number of publications per author are the bipartite degrees, while the others are taken from the projections. In the *weighted projection* each contains the number of *shared bipartite* neighbours, e.g. the edge weight between two authors carries the number of co-authored publications. The *degree* describes the size of the neighbourhood of a node, while the *strength* also takes those edge weights into account. The degree and strength distributions are strongly right skewed, and look similar to the ubiquitous scale-free networks (Barabasi 2002), but the data set is too small to proof scale-freeness.

About one third (authors), and more than half (publications) of the nodes belong to the largest component. There are 334 components of authors (publications) which are not connected to each other - at least not inside the ISI-UCD dataset. In both projections we see a sparse graph. The clustering coefficient is high - the mean probability that collaborators of an author have published a common paper themselves is 18% in UCD, which is lower than e.g. the 30-40% found in extensive physics publication databases (Newman, 2001) due to the restriction to UCD and ISI (omitting other links between authors).

A "mean UCD-author" has published around 15 papers mentioned in the ISI index, with the most prolific author having published 218 papers. The greatest number of co-authorships is 63, far above the average of 3 co-publications for collaborating pairs of authors at UCD. The most connected author has 18 UCD co-authors, in comparison to the mean of just 2 co-authors. All distributions are strongly right-skewed; most values are small, but very large values also appear - so all given mean values are to be treated with cautio

| **network measure** | **explanation** | **author projection**<br>author -(publication)- author | **publication projection**<br>publication -(author)- publication |
|---|---|---|---|
| vertex type: | basic entity in projection | author | publication |
| edge type: | condition for linking | co-authored publication | same author |
| #vertices N | size of population | 642 | 7,911 |
| #edges M | total number of links | 635 | 178,972 |
| edge density ρ | ρ=M / (N*(N-1)/2) | 0.0031 = 1/324.0 | 0.0057 = 1/174.8 |
| max #triangles | in neighbourhood of vertex | 42 | 23436 |
| mean #triangles | calc'ed over all vertices | 1.83 | 1810.2 |
| clustering coefficient | #triangles / #possible triangles, mean over all vertices | 0.18 | 0.94 |
| #components | # unconnected partitions | 334 | 334 |
| l.c. #vertices | largest component (% of N) | 234 (36%) | 4,395 (56%) |
| l.c. #edges | largest component (% of M) | 540 (85%) | 113,007 (63%) |
| l.c. edge density | largest component | 0.0198 = 1/50.5 | 0.0117 = 1/85.4 |
| l.c. diameter | longest geodesic path | 12 | 13 |
| l.c. mean pathlength | of all geodesic pathlengths | 4.69 | 5.05 |

Table A1: Standard unweighted network measures for the two projections.

| distribution | | mean | st.dev | skewness | max |
|---|---|---|---|---|---|
| # UCD-authors per publication | omitting non-UCD authors | 1.19 | 0.51 | 3.11 | 5 |
| publication projection | degrees (unweighted) | 45.3 | 43.9 | 2.02 | 226 |
| | edge weights = # identical authors of 2 publications | 1.05 | 0.23 | 5.41 | 4 |
| | strengths (weighted degree) | 47.8 | 47.2 | 2.03 | 325 |
| # publications per UCD-author | represented in ISI database | 14.7 | 19.6 | 3.61 | 218 |
| author projection | degrees (unweighted) | 2.0 | 3.16 | 4.56 | 18 |
| | edge weights # repeated co-authoring | 3.04 | 4.67 | 5.88 | 63 |
| | strengths (weighted degree) | 11.3 | 16.0 | 3.79 | 128 |

Table A2: Properties of distributions derived from the network edges.

**Acknowledgement**
This paper is based on research sponsored by UCD Research and by the European FP6-NEST-Adventure Programme (contract number 028875). Portions of the work were done at the Center for Mathematical Sciences at the University of Madeira during Madeira Math Encounters XXXVI. Thanks also to Andrej Mrvnar for his assistance with Pajek.

## References

Barabasi, A. L., Jeong, H., Neda, Z., Ravasz, E., Schubert, A., Vicsek, T., (2002). Evolution of the social network of scientific collaborations, Physica A: Statistical Mechanics and its Applications, Volume 311, Issues 3-4, 15, 590-614

Barabasi, A.L., Albert, R. (2002) Statistical mechanics of complex networks, Reviews of Modern Physics 74, 47, arxiv:cond-mat/0106096

Barber, M. J. (2007). Modularity and community detection in bipartite networks. *Physical Review E (Statistical, Nonlinear, and Soft Matter Physics)*, 76(6):066102.

Bozeman, B., Lee, S. (2005). The Impact of Research Collaboration on Scientific Productivity, Social Science Studies, 35 (5): 673-702

Burt, R. S. (2004) Structural Holes and Good Ideas, The American Journal of Sociology, Vol. 110, No. 2. pp. 349-399

Clauset, A., Newman, M. E. J., and Moore, C. (2004). Finding community structure in very large networks. *Physical Review E (Statistical, Nonlinear, and Soft Matter Physics)*, 70(6):066111.

Danon, L., Dìaz-Guilera, A., Duch, J., and Arenas, A. (2005). Comparing community structure identification. *J. Stat. Mech.*, page P09008.

Edquist, C., (1997). Systems of Innovation: Technologies, Institutions and Organizations, Pinter publishers, New York/London.

Etzkowitz, H., Leydesdorff, L., (2000). The dynamics of innovation: from National Systems and “Mode 2” to a Triple Helix ofuniversity–industry–government relations. Research Policy 29 (2),109–123.

Fleming, Lee, Santiago Mingo, and David Chen. (2007) Collaborative Brokerage, Generative Creativity, and Creative Success, Administrative Science Quarterly 52, no. 3

Fortunato, S. and Castellano, C. (2008). Community structure in graphs. In *Encyclopedia of Complexity and System Science*. Springer.

Jones, B F. Wuchty, S, Uzzi, B. (21 November 2008) Multi-University Research Teams: Shifting Impact, Geography, and Stratification in Science, Science 322 (5905), 1259.

Gibbons, M., Limoges,C., Nowotny, H., Schwartzman, S., Scott, P., Trow, M.,(1994). The New Production of Knowledge: The Dynamics of Science and Research in Contemporary Societies. SAGE, London.

Gloor, P. (2006) Swarm Creativity: Competitive Advantage Through Collaborative Innovation Networks. Oxfard University Press

Guimerà, R., Uzzi, B., Spiro, J., Nunes Amaral, L. (2005). Team Assembly Mechanisms Determine Collaboration Network Structure and Team PerformancE, Science 308 (5722), 697.

Heinze,T. Shapira,P. Rogers, J.D.,Senker, J.(2009) Organizational and institutional influences on creativity in scientific research, Research Policy, Volume 38, Issue 4,610-623

Katz, J. Sylvan & Martin, Ben R. (1997). What is research collaboration? Research Policy, Elsevier, vol. 26(1), 1-18.

Katz, S, Hicks, D, (1997). How much is a collaboration worth? A calibrated bibliometric model. Proceedings on the Sixth Conference of The International Society for Scientometric and Informetric, Jerusalem, Israel, June 16-19, 163-175.

Laudel, G. (2002). What do we measure by co-authorships?, Research Evaluation, 11 pages, 3-15.

Laudel, G, Gläser, J, (2006). Tensions between evaluations and communication practices. Journal of Higher Education Policy & Management; Vol. 28 Issue 3, 289-295

Newman, M. E. J.(2001) Who is the best connected scientist? A study of scientific coauthorship networks, Phys.Rev. E64 016131; arXiv:cond-mat/0011144v2

Newman, M. E. J. (2004a). Detecting community structure in networks. *Eur. Phys. J. B*, 38:321–330.

Newman, M. E. J. (2004b). Fast algorithm for detecting community structure in networks. *Physica l Review E (Statistical, Nonlinear, and Soft Matter Physics)*, 69(6):066133.

Newman, M. E. J. and Girvan, M. (2004). Finding and evaluating community structure in networks. *Physical Review E (Statistical, Nonlinear, and Soft Matter Physics)*, 69(2):026113.

Olmeda-Gómez, Perianes-Rodríguez, Ovalle-Perandones, Moya-Amegon, (2008). Comparative analysis of university-government-enterprise co-authorship networks in three scientific domains in the region of Madrid. Information research, vol. 13 no. 3

Slaughter, S., Leslie, L. , (1999). Academic capitalism: Politics, policies, and the entrepreneurial university. Baltimore: Johns Hokpins.

Uzzi, B., Spiro, J., 2005. Collaboration and creativity: the small world problem. American Journal of Sociology 111, 447–504.

van Raan, Anthony F. J.(2005). Measurement of Central Aspects of Scientific Research: Performance, Interdisciplinarity, Structure',Measurement: Interdisciplinary Research & Perspective,3:1,1-19.

Wasserman, S. and Faust, K. (1994) Social Network Analysis: Methods and Applications. Cambridge: Cambridge University Press.

Wuchty,S. Jones, B. Uzzi, B, (2007). The Increasing Dominance of Teams in Production of Knowledge. Science Express, Vol. 316. no. 5827, pp. 1036 - 1039

Yousefi-Nooraie, R, Akbari-Kamrani, M , Hanneman, R, Etemadi, (2008). Association between co-authorship network and scientific productivity and impact indicators in academic medical research centers: A case study in Iran, Health Research Policy and Systems 2008, 6:9.

Ziman, J., (2000). Real Science: What it is, and What it Means. Cambridge University Press, Cambridge.